\documentclass{PoS}
\usepackage{float}
\title{Modeling the elastic differential cross-section at LHC}
\ShortTitle{Modeling the elastic differential cross-section at LHC}

\author{\speaker{D. A. Fagundes}\\
INFN Frascati National Laboratories, Via E. Fermi 40, 00444, Italy\\
Instituto de F\'{\i}sica Gleb Wataghin, Universidade Estadual de Campinas, UNICAMP, 13083-859 Campinas SP, Brazil\\
E-mail: \email{fagundes@ifi.unicamp.br}}

\author{A. Grau\\
Departamento de F\'{\i}sica Te\'orica y del Cosmos, Universidad de Granada, 18071 Granada, Spain\\
E-mail: \email{igrau@ugr.es}}

\author{S. Pacetti\\
Physics Department and INFN, University of Perugia, 06123 Perugia, Italy\\
E-mail: \email{simone.pacetti@pg.infn.it}}

\author{G. Pancheri\\
INFN Frascati National Laboratories, Via E. Fermi 40, 00444, Italy\\
E-mail: \email{giulia.pancheri@lnf.infn.it}}

\author{Y. N. Srivastava\\
Physics Department and INFN, University of Perugia, 06123 Perugia, Italy\\
E-mail: \email{yogendra.srivastava@pg.infn.it}}

\FullConference{XXI International Workshop on Deep-Inelastic Scattering and Related Subject -DIS2013,\\ 22-26 April 2013\\ Marseille, France}

\abstract{An empirical model for the $pp$ elastic differential cross section is proposed. Inspired by early work by Barger and Phillips, we parametrize the scattering amplitude in building blocks, comprising of two exponentials with a relative phase, supplementing the dominant term at small $-t$ with the proton form factor. This model suitably applies to LHC7 and ISR data, enabling to make simple predictions for higher LHC energies and to check whether asymptotia might be achieved.
}

\begin{document}

\section{Introduction} 
This contribution aims at examining the main features of one of the relevant observables for particle diffraction at high-energies, namely the elastic differential cross section. We investigate the structure of the recent TOTEM data for the LHC run at $\sqrt{s} = 7 $ TeV (LHC7) \cite{lhc7data}: the \textit{diffraction cone}, the sharp `dip' struture and the large $-t$ region. By means of an \textit{empirical} parametrization, based on the Barger-Phillips model (henceforth called BP) \cite{bp,gpps} we perform fits to the present data on $pp$ scattering, analysing its applicability in the wide energy range from 24 GeV to 7 TeV. While keeping the original structure of the BP model, namely the two building blocks comprising of two exponential terms 
interfering through a relative phase, we propose to use the following modified version 
\vspace*{-0.1cm}
\begin{eqnarray}
\mathcal{A}(s,t)=i[F^2_P(t)\sqrt{A(s)}e^{B(s)t/2}+e^{i\phi(s)} \sqrt{C(s)}e^{D(s)t/2}], \label{eq:dis01}
\end{eqnarray}
with the first term supplemented by $F_{p}^{2}(t) = 1/(1-t/t_{0})^4$, the proton form factor. As was the case  for the original BP amplitude, this amplitude can be interpreted at the light of contributions with opposite parities, $C=\pm 1$. The $\sqrt{A}-$term, being the leading contribution at small $-t$, might be related to a $C=+1$ exchange, while the $\sqrt{C}-$term, comprising the phase $\phi$, encompass even and odd parities and is considered nonleading. This parametrization has been recently discussed in detail in Ref. \cite{fgpps}, where it was found that the correction, introduced by the form factor to the leading term of the BP amplitude, improves  the description of data at very small $-t$ values. 
Admittedly while this is not the only option, we outline here three main reasons to focus on this particular model: (i) it describes very well the LHC7 and ISR data; (ii) the form factor adds physical content to the BP amplitude - being the probability that the proton does not break up in the collision; (iii)  it allows simple predictions for higher LHC energies.\\
In the following, after presenting our fit results, we discuss the method applied to obtain the energy depence of fit parameters in the model (\ref{eq:dis01}). This will then allow us to make predictions for future LHC energies and to test asymptotia in $pp$ collisions. 
\section{Fit with the modified BP model}
\begin{figure}[h]
\centering
\includegraphics*[width=7.5cm,height=7.0cm]{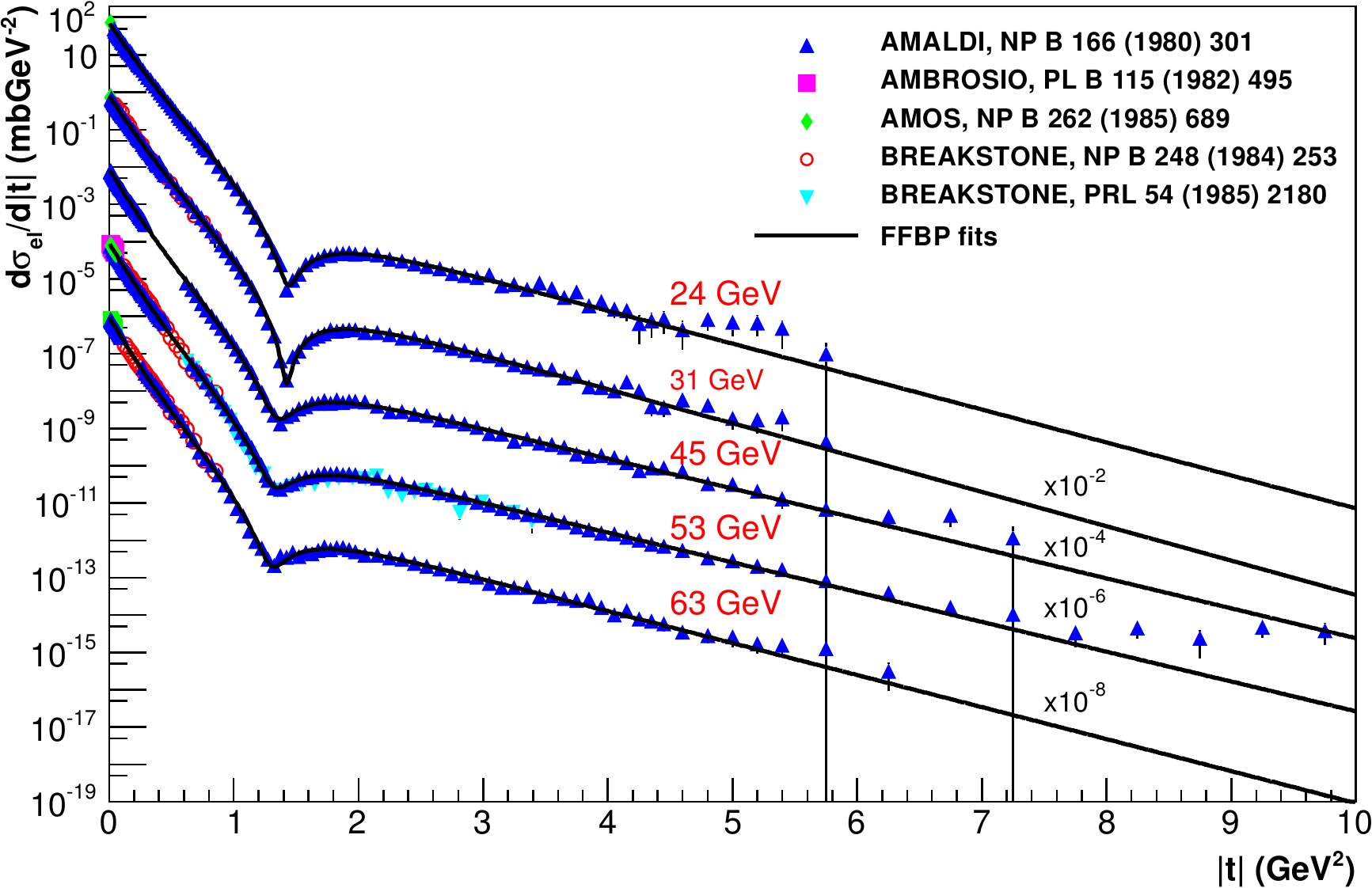}\hspace*{0.2cm}
\includegraphics*[width=7.5cm,height=7.0cm]{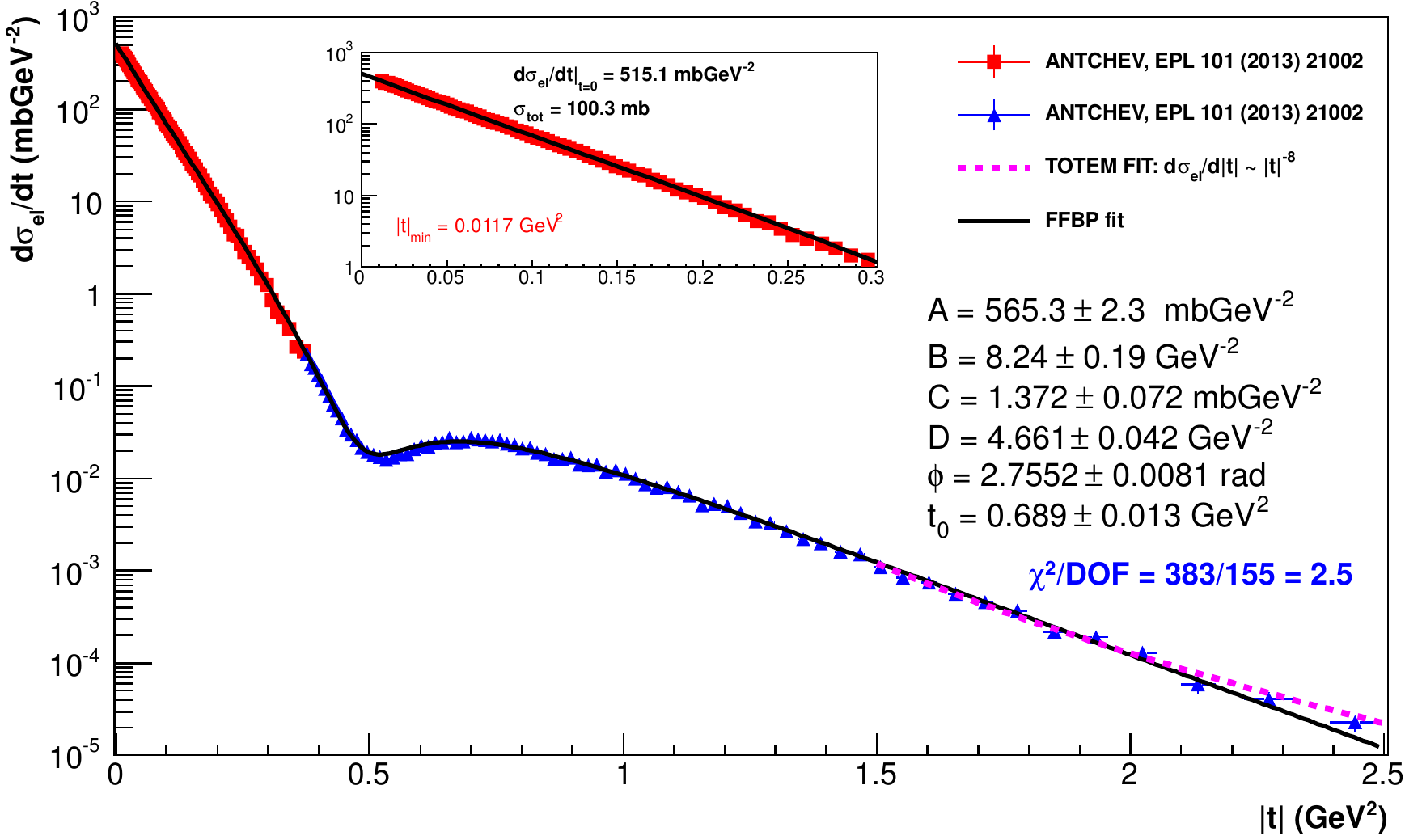}
\caption{Fits to the ISR and LHC7 data sets with modified BP model of Eq.(\protect\ref{eq:dis01}).}
\label{fig:dis01}
\end{figure}

In Fig. \ref{fig:dis01} and Table \ref{tab:dis01} we summarize our fit results with the model (\ref{eq:dis01}). ISR data were compiled from \cite{isrdata}. These results provide evidence for the fact that at LHC7, the scale $t_{0}$, present in the proton form factor, is consistent with the EM one, 0.71 GeV$^{2}$, and may be fixed at this value without worsening the quality of fit (see the last two rows of Table \ref{tab:dis01}). 
In effect, the 
monotonic decrease of $t_{0}(s)$ from ISR to LHC7 led us to hypothesize that for $\sqrt{s} \geqslant $ 7 TeV this parameter 
saturates at 0.71 GeV$^{2}$ and  on making predictions for higher energies with this model, we assume  $t_{0} $ asymptotes to 0.71 GeV$^{2}$. 
 
\vspace*{-0.3cm}
\begin{table}[H]
\caption{
Free fit parameters $A, B, C, D, t_{0}$ and $\phi$ of the model (\protect\ref{eq:dis01}) at each energy analyzed. 
In the last row, the scale parameter $t_0$ is kept fixed. $A$ and $C$ are expressed in units of mbGeV$^{-2}$, $B$ 
and $D$ in units of GeV$^{-2}$, $t_{0}$ in units of GeV$^{2}$, and $\phi$  in radians.} 
\vspace{0.2cm}
\centering
{\scriptsize
\begin{tabular}{|c|c|c|c|c|c|c||c|c|}
\hline \hline 
$\sqrt{s}$ (GeV) & $A$ & $B$ & $C (\times 10^{-3})$  & $D$  & $t_{0}$  & $\phi$ & $DOF$ &$\frac{\chi^{2}}{\rm DOF}$\\ 
\hline 24 &  74.8 $\!\pm\!$ 0.8 & 4.0 $\!\pm\!$ 0.1 &  4.8 $\!\pm\!$ 0.7  &  2.03 $\!\pm\!$ 0.06
  & 1.06 $\!\pm\!$ 0.03  & 3.31 $\!\pm\!$ 0.01  & 128 & 1.2 \\  
\hline 31 & 83.7 $\!\pm\!$ 0.2   &  3.90 $\!\pm\!$ 0.07  &  5.4 $\!\pm\!$ 0.5  & 2.12 $\!\pm\!$ 0.04 
&  0.99 $\!\pm\!$ 0.01  & 3.06 $\!\pm\!$ 0.01  & 200 & 1.6 \\ 
\hline 45 & 89.6 $\!\pm\!$ 0.2   &  4.27 $\!\pm\!$ 0.05  & 2.4 $\!\pm\!$ 0.2  &  1.84 $\!\pm\!$ 0.02
& 0.912 $\!\pm\!$ 0.009   & 2.83 $\!\pm\!$ 0.01 & 201 & 3.7 \\  
\hline 53 & 93.0  $\!\pm\!$ 0.1   & 4.51 $\!\pm\!$ 0.05  & 2.5 $\!\pm\!$ 0.1   & 1.84 $\!\pm\!$ 0.01   & 0.947 $\!\pm\!$ 0.008  & 2.79 $\!\pm\!$ 0.01   & 313 & 4.7 \\ 
\hline 63 & 97.4  $\!\pm\!$ 0.2   & 4.3 $\!\pm\!$ 0.1   & 3.5 $\!\pm\!$ 0.4   & 1.97 $\!\pm\!$ 0.04  & 0.90 $\!\pm\!$ 0.01 &  2.86 $\!\pm\!$ 0.06   & 159 & 2.1 \\
\hline 7000 & 565 $\!\pm\!$ 2  & 8.2 $\!\pm\!$ 0.2 &  1370 $\!\pm\!$ 70   & 4.66 $\!\pm\!$ 0.04 & 0.69 $\!\pm\!$ 0.01 & 2.755 $\!\pm\!$ 0.008 & 155 & 2.5\\
\hline 7000 & 562 $\!\pm\!$ 1  & 8.54 $\!\pm\!$ 0.03 &  1280 $\!\pm\!$ 34   & 4.61 $\!\pm\!$ 0.03 & 0.71 (fixed)  & 2.744 $\!\pm\!$ 0.004 & 156 & 2.5\\
\hline \hline
\end{tabular}}
\label{tab:dis01}
\end{table} 

\section{Asymptotic sum rules}
The BP model provides a suitable framework to check asymptotia in $pp$ collisions \cite{gpps}, through the application of two asymptotic sum rules in the impact parameter space. In the scope of the BP parametrization, they follow straightforwardly:
\begin{eqnarray}
SR_1 = \frac{1}{\sqrt{\pi}} [\frac{ \sqrt{(\frac{A}{1 + {\hat \rho}^2})} }{B} - \frac{\sqrt{C}}{D} |\cos \phi|]; 
\ \ \ \ 
SR_0= \frac{1}{\sqrt{\pi}} [\frac{ \sqrt{(\frac{A}{1 + {\hat \rho}^2})} }{B} \hat{\rho} - \frac{\sqrt{C}}{D} \sin \phi]; \label{eq:dis03}
\end{eqnarray}
with $\hat{\rho}$ being the contribution to real part of the amplitude, originated from the first term. Since we analyze here a somewhat different model, 
Eq. (\ref{eq:dis03}) does not exactly correspond to the sum rules obtained from Eq. (\ref{eq:dis01}). Even though the new parametrization does introduce some changes - mainly due to the presence of the form factor $F^{2}_{p}(t)$ - we argue that it is not going to spoil these simple relations. In effect, a very similar expression for these sum rules was derived in \cite{fgpps}, taking $t_{0}$ into account. Ultimately, their satisfaction at asymptotic energy leads to:
\begin{equation}
\label{eq:dis04}
SR_1 \to\ 1- ;\ SR_0 \to\ 0+.
\end{equation} 
Using the fit parameters in Table \ref{tab:dis01} we have calculated $SR_{1}$ and $SR_{0}$, as they actually follow from Eq. (\ref{eq:dis01}). A summary of these results is thus shown in  Table \ref{tab:dis02}.     

\begin{table}[H]
\caption{Sum rules for the modified BP model (\protect\ref{eq:dis01}) at two ISR energies (24 GeV and 53 GeV) and at LHC7. For the calculations we have adopted an specific model for $\hat{\rho}(s)$ \cite{bnmodel} - $\hat{\rho}(s)=\pi/2p\ln s$, with $p$ constrained in the interval $1/2<p<1$.}
\centering
\vspace*{0.2cm}
\begin{tabular}{c|c|c|c}
\hline\hline  $p$ & $\sqrt{s}$ (GeV) & $SR_{1}$ & $SR_{0}$ \\ 
\hline
$-$ & 24 & 0.719 & 0.021 \\
$-$ &  53 &  0.717 & 0.049 \\
0.66 &  7000 &  0.950 & 0.070 \\
0.77 & 7000 & 0.953 & 0.048 \\
\hline\hline
\end{tabular} 
\label{tab:dis02}
\end{table}
Given these results, we see  that, when compared with the original parametrization, the modified BP model improves the satisfaction of the sum rules. Although, at present energies \textit{true} asymptotia is not yet realized, the above results suggest that we are approaching the bounds (\ref{eq:dis04}). Assuming their saturation at higher energies, next we propose to make predictions for the energy behaviour of the parameters of the model in context.

\section{Energy evolution of parameters and predictions for LHC8 and LHC14}
\begin{figure}[H]
\centering
\includegraphics*[width=7.7cm,height=7.5cm]{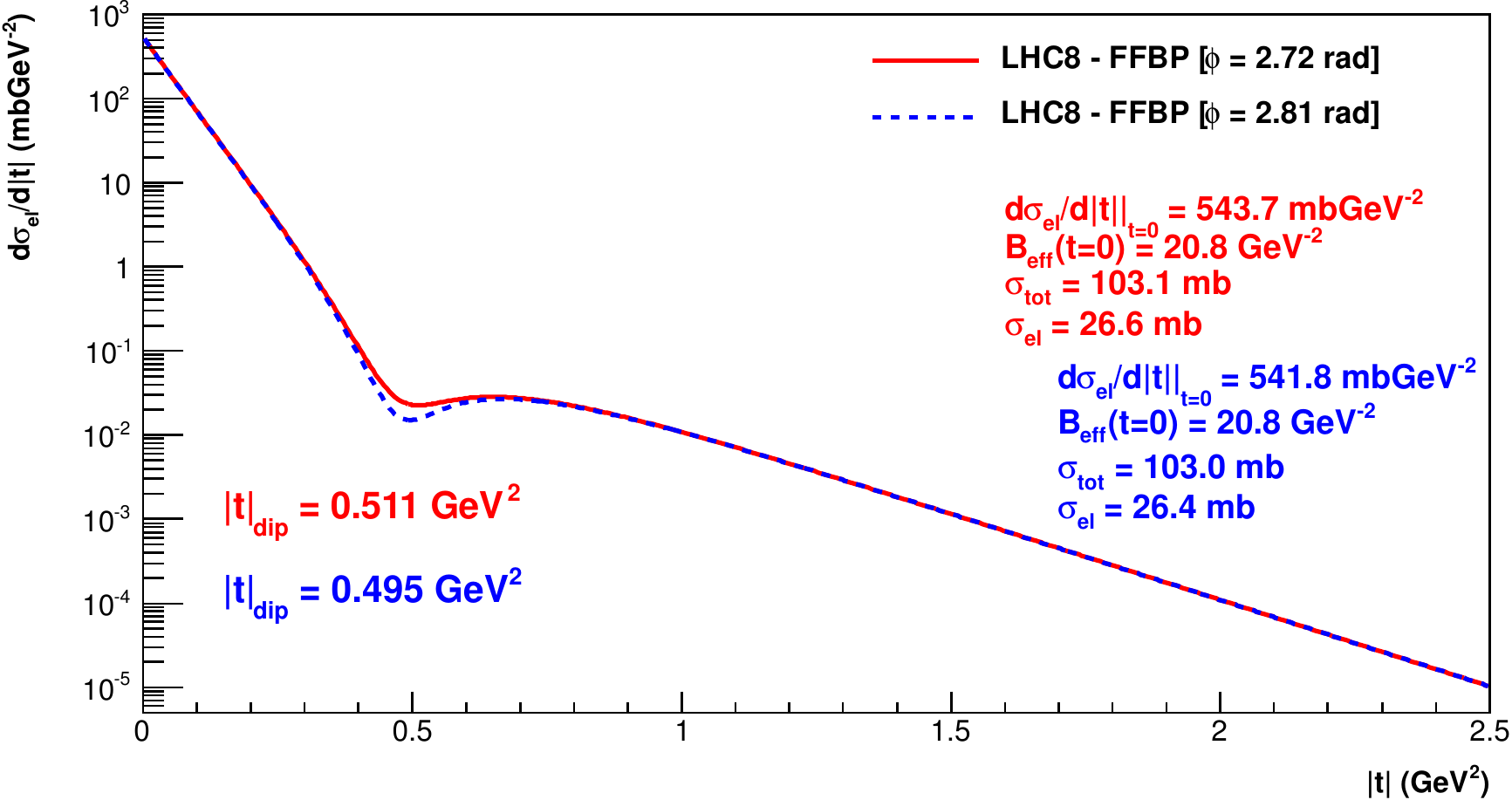}\hspace*{0.05cm}
\includegraphics*[width=7.7cm,height=7.5cm]{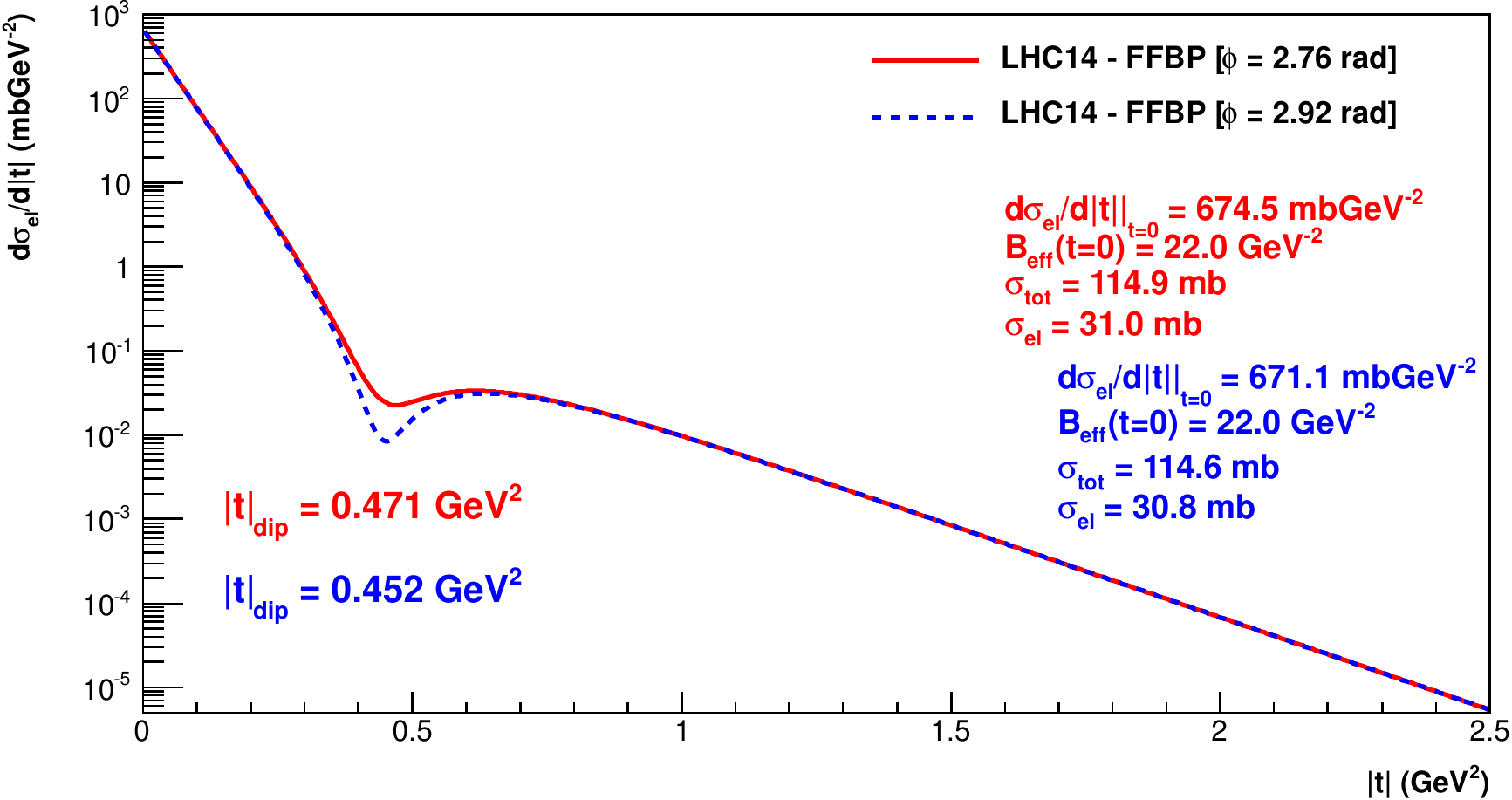}
\caption{Predictions for the differential elastic cross section at LHC8 and LHC14 from the modified BP model (\protect\ref{eq:dis01}), in a maximal energy saturation regime, where $\sigma_{total} \sim (\ln s)^2$.}
\label{fig:dis02}
\end{figure}

While our fit results do not allow a complete determination of the energy dependence of all fit parameters, they support: (i) $t_{0}\rightarrow$ 0.71 GeV$^{2}$, for $\sqrt{s}\geqslant 7$ TeV; (ii) $\phi \sim constant$ over a wide energy range - spanning from ISR24 to LHC7. Thus, under the reasonable assumption that asymptotically  $t_{0}$ and $\phi$ will become constant, and having $\hat{\rho} \sim 1/\ln s$ (as required by the Khuri-Kinoshita theorem \cite{kk}) we obtain the following relationship between parameters:

\begin{eqnarray}
\frac{\sqrt{A(s)}}{B(s)}&\sim& \frac{\sqrt{C(s)}}{D(s)}\ln s;\label{eq:dis05}\\
\frac{\sqrt{A(s)}}{B(s)}&\sim& \frac{\sqrt{\pi}}{(1+\frac{\pi \cot \phi}{2p\ln s})} \sim constant.\label{eq:dis06}
\end{eqnarray} 
To derive the latter we took $\hat{\rho}(s)=\pi/2p\ln s$ from the model \cite{bnmodel}. On the one hand, from the amplitude (\ref{eq:dis01}), the connection with the optical point leads to: $A(s) \propto \sigma_{tot}^{2}$. On the other, Eq. (\ref{eq:dis06}) expresses the asymptotic equivalence: $\sigma_{tot} \sim B(s)$. \\
Here we will consider the particular case of \textit{maximal} energy behaviour allowed by Froissart-Martin bound \cite{fmbound}. Therefore, in this scenario it follows from Eq. (\ref{eq:dis05}) that:
\begin{eqnarray}
\sqrt{A(s)} \sim \ln^{2} s, \quad B(s) \sim \ln^{2} s, \quad D(s) \sim  \sqrt{C(s)}\ln s.\label{eq:dis07}
\end{eqnarray}
For simplicity, we restrict our analysis to the \textit{plausible} case of Regge-like behaviour  of the slope $D(s)$ (albeit not the only one), appearing in the nonleading term of Eq. (\ref{eq:dis01}). Finally, we propose the asymptotic solutions
\begin{eqnarray}
\sqrt{A(s)} \sim \ln^{2} s, \quad B(s) \sim \ln^{2} s, \quad D(s) \sim \ln s, \quad \quad \sqrt{C(s)} \sim constant.\label{eq:dis08}
\end{eqnarray}  
Our parametrizations for each of these parameters, in the energy range analyzed, follow below:
\begin{eqnarray}
4  \sqrt{\pi A(s)} &=& 47.8-3.8 \ln s+0.398 (\ln s)^2\ [mb]; \label{eq:dis09}\\
B(s) &=& -0.23+0.028 (\ln s)^2\ [GeV^{-2}] ;\label{eq:dis10}\\
4 \sqrt{\pi C(s)} &=& \frac{9.6 -1.8 \ln s +0.01( \ln s)^3}{1.2+0.001(\ln s)^3}\ [mb];\label{eq:dis11}\\
D(s) &=& -0.41+0.29 \ln s\ [GeV^{-2}]. \label{eq:dis12}
\end{eqnarray}
The energy dependence of the amplitude $A(s)$ and the slopes $B(s)$ and $D(s)$ were extracted from fits to the data shown in Table \ref{tab:dis01} and are motivated by the asymptotic behaviour given in Eqs. (\ref{eq:dis08}). For $C(s)$ an \textit{empirical} formula is given, satisfying the asymptotic condition $\sqrt{C(s)}\sim constant$. Keeping $t_{0}=0.71$ GeV$^{2}$ and $\phi \simeq 2.7-2.9\ rad$, as suggested by our phenomenology, enables us to get predictions for LHC8 and LHC14, using Eqs. (\ref{eq:dis09}-\ref{eq:dis12}), as we show in Fig. \ref{fig:dis02}. Ultimately, our lack of knowledge about the phase $\phi$ prevents a precise determination of its value at higher energies. But on holding on to the hypothesis of $\phi \sim constant$ at higher energies, two predictions are given for the elastic differential cross section at LHC8 and LHC14. The crucial role of $\phi$  
in determining the diffractive minimum, namely its position and depth, is easily seen from these plots.
As we show next, even though essential to specify the position of the dip, the actual value of $\phi$ is less relevant for the integrated cross sections -   consequently for the ratio $R_{el}(s)=\sigma_{el}(s)/\sigma_{tot}(s)$, for which predictions with different values of $\phi$ practically overlap. The asymptotic predictions of model (\ref{eq:dis01}) for the ratio $R_{el}(s)$ allow simple tests of where true asymptotia in $pp$ scattering might be reached.

\section{The black disk limit}
\begin{figure}[H]
\begin{center}
\includegraphics[width=13cm,height=8.0cm]{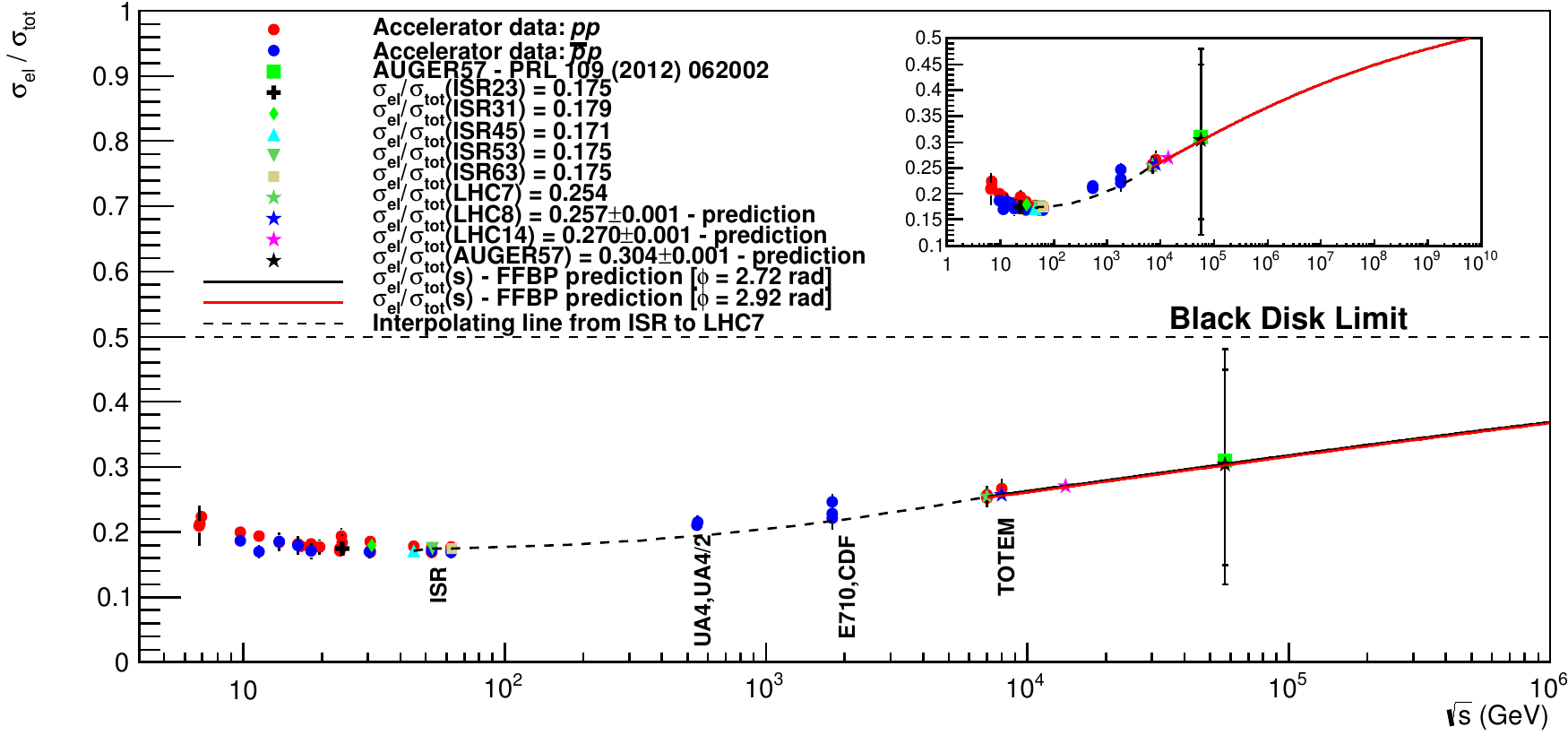}
\caption{Experimental data of the ratio $R_{el}=\sigma_{elastic}/\sigma_{total}$ and predictions from the model (\protect\ref{eq:dis01}) in the energy region $\sqrt{s}> 7$ TeV. The AUGER datum at $\sqrt{s} = 57$ TeV was estimated from the results given in \cite{augerdatum} - \textit{inner bars} comprise only statistical and systematic uncertainties combined quadratically, while \textit{outer bars} incorporate the total uncertainty, with errors from Glauber calculations also summed in quadrature.
Inner bars:  $R_{el}^{stat+sys}(57 TeV) = 0.31^{+0.14}_{-0.16}$,
outer bars:  $R_{el}^{stat+sys+Glauber}(57 TeV) = 0.31^{+0.17}_{-0.19}$.}
\label{fig:rel}
\end{center}
\end{figure}

The asymptotic satisfaction of the sum rules, reinforcing the condition of total absorption of partial waves, leads to the saturation of the black disk limit, i.e. $R_{el}\rightarrow 1/2$ as $s\rightarrow \infty$. For the energy region $\sqrt{s}> 7$ TeV, numerical extrapolations to the energy frontier needed to achieve this limit are given in Fig. \ref{fig:rel}. With our model we find: $R_{el} \simeq 1/2$ at $\sqrt{s}\simeq 10^{10}$ GeV (corresponding to $E_{lab}\simeq 10^{20}$ GeV) - an energy typically larger than the Planck scale and really far from being reached.

\end{document}